\documentclass[twocolumn,showpacs,preprintnumbers,amsmath,amssymb,prl]{revtex4}
\usepackage{graphicx}
\usepackage{dcolumn}
\usepackage{bm}
\usepackage{amsmath}
\usepackage{subfigure}
\begin{document}
\title{Integrated photonic qubit quantum computing on a
  superconducting chip} \author{Lianghui Du, Yong Hu, Zheng-Wei Zhou,
  Guang-Can Guo and Xingxiang Zhou} \email{xizhou@yahoo.com}
\affiliation{Key Laboratory of Quantum Information and Department of
  Physics, University of Science and Technology of China, Chinese
  Academy of Sciences, Hefei, Anhui 230026, China } \date{\today}

\begin{abstract}
  We study a quantum computing system using microwave photons in
  transmission line resonators on a superconducting chip as qubits. We
  show that all control necessary for quantum computing can be
  implemented by coupling to Josephson devices on the same chip. We
  take advantage of the strong nonlinearities inherent in Josephson
  junctions to realize qubit interactions. We analyze the gate error
  rate to demonstrate that our scheme is realistic even for Josephson
  devices with limited decoherence times. A conceptually innovative
  solution based on existing technologies, our scheme provides an
  integrated and scalable approach to the next key milestone for
  photonic qubit quantum computing.

\end{abstract}


\maketitle

Despite the vast potential of quantum computers, no perfect physical
implementation has been found for quantum computers. This can be seen
by examining two representative systems. Josephson device based
superconducting systems are easily integrable and scalable, but are
plagued by the short decoherence times of Josephson qubits due to
coupling to their complex solid-state environment. Photonic qubits,
which have superb coherence properties, suffer from the fact that
photons do not interact easily. Also, systems based on conventional
bulk optical devices are hard to miniaturize and scale.

Recognizing the importance of integrated systems for scalable quantum
computing, a number of investigators have demonstrated on-chip
waveguide based quantum gates for photonic qubits recently
\cite{ref:Integrated_OQC}. This is a significant step that may
represent the future direction of photonic qubit quantum computing
technologies. However, it is a daunting task to achieve a fully
integrated photonic qubit quantum computer using conventional
technologies including those developed in the latest experiments.
This is because conventional optical devices such as lasers, lenses,
optical cavities and photo detectors are bulk devices based on very
different technologies and no process exists yet to integrate them on
the same chip \cite{ref:onchip}. Therefore, alternative realistic
approaches to fully integrated photonic qubit quantum computing
systems are highly valuable.

We combine the strengths of photonic and superconducting systems to
realize fully integrated photonic qubit quantum computing. Our
physical system is a superconducting chip on which high-Q transmission
line resonators (TLRs) and Josephson devices are fabricated. The same
system has been used for study of cavity QED based on Josephson qubits
\cite{ref:CJ, ref:tlr, ref:disp}. However, in our scheme the quantum
information is carried by the microwave photon modes in the TLRs and
the Josephson junctions play the role of optical devices. For high-Q
TLRs the photons have a long life time \cite{ref:highQ} which is a
major advantage. Easy operation and accurate control are available
because Josephson devices can be fabricated with great precision and
controlled conveniently by monitoring their electrical signals. A
further key advantage is we can use the strong nonlinearities inherent
in Josephson devices to induce interactions between photons. It is
shown that high gate fidelities can be achieved even for Josephson
devices with limited decoherence times making their unavoidable noisy
environment no longer a limiting factor. Therefore, our scheme is a
realistic approach to scalable photonic qubit quantum computing.

\begin{figure}[h]
\subfigure []
{\label{1a}\includegraphics[width=40mm, height=25mm]{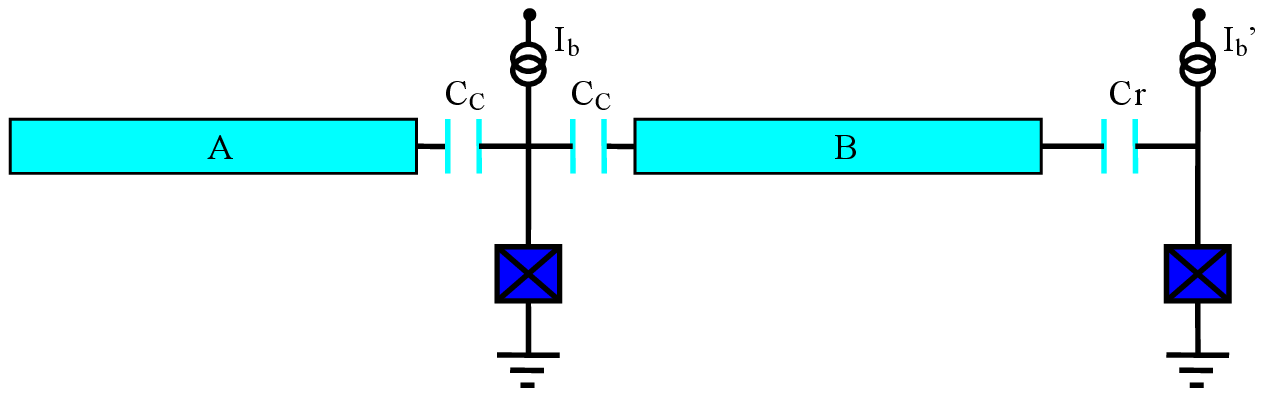}}%
\subfigure[]{\label{1b}\includegraphics[width=40mm, height
=40mm]{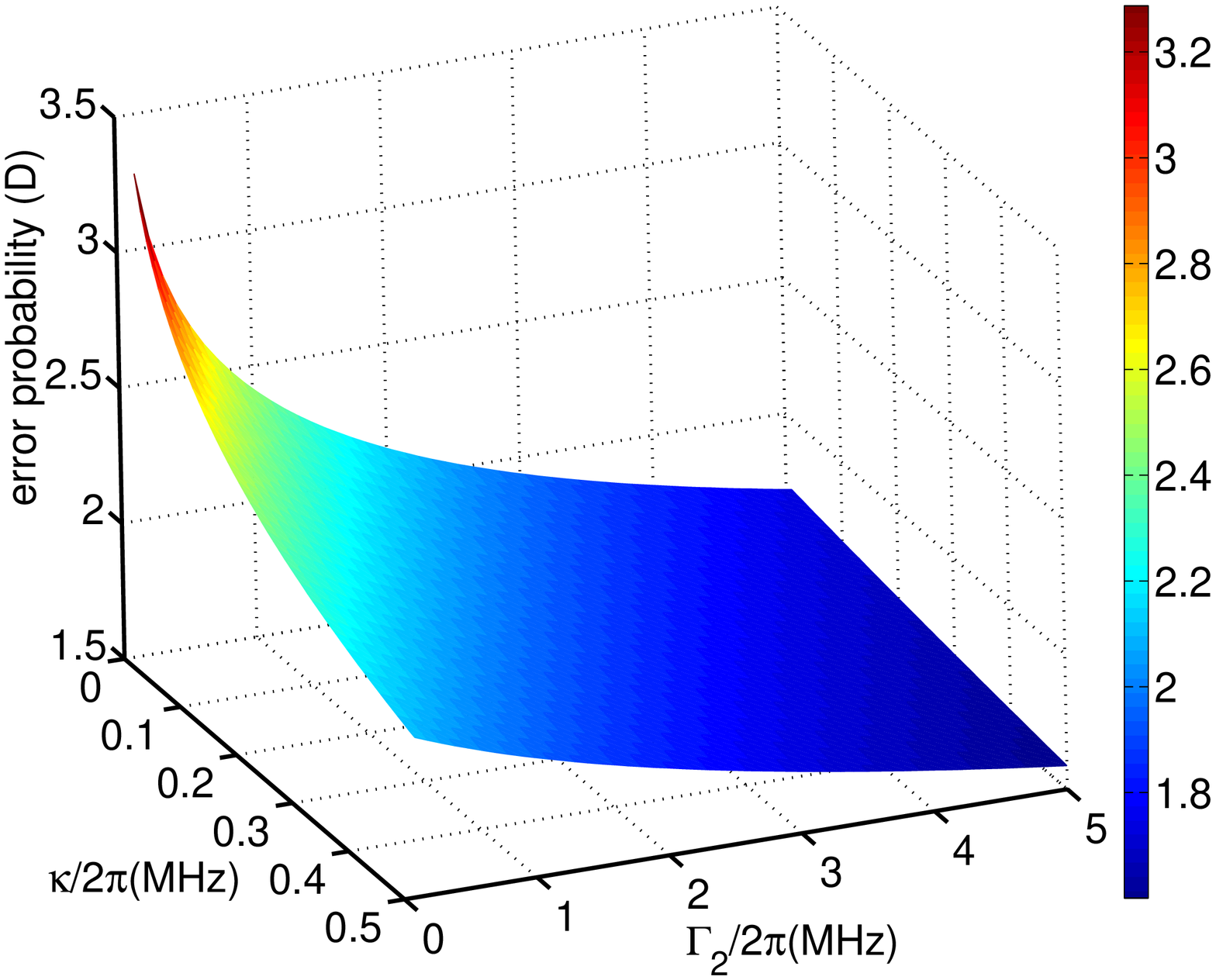}} \caption{(a) A photonic qubit based on two TLRs.
$A$ and $B$ are capacitively coupled to the coupling CBJJ via $C_c$.
$B$ is also coupled to the right CBJJ via $C_r$. (b) Error
probability ($10^{-D}$) in realizing the photon transfer operation
between $A$ and $B$ as a function of the photon loss rate $\kappa$
and CBJJ dephasing rate $\Gamma_2$.}\label{fig1}
\end{figure}

We start by considering the two identical TLRs shown in
Fig.\ref{fig1}(a). The TLR mode frequencies are given by
$\omega=n\pi/\sqrt{LC}$, $n$ an integer and $L$ and $C$ the total
inductance and capacitance of the TLR. We use the $n = 2$ mode. For $L
= 0.5$nH and $C = 5$pF, its frequency $\omega_0/2\pi \approx
20$GHz. The second-quantized voltage and current associated with this
mode is $V(x, t)=\sqrt{\hbar\omega_0/C} \cos{\frac{2\pi
    x}{l}}(\hat{a}(t) + \hat{a}^{\dagger}(t))$ and $I(x, t)= -i
\sqrt{\hbar\omega_0/L} \sin{\frac{2\pi x}{l}}(\hat{a}(t) -
\hat{a}^{\dagger}(t))$, where $l$ the length of the TLR, $x \in
[-l/2,l/2]$ the position along the TLR, and $\hat{a}(t) =
\hat{a}e^{-i\omega_0 t}$ the mode's annihilation operator.

For the pair of identical TLRs in Fig.\ref{fig1}(a), we introduce a
single photon of frequency $\omega_0$ and it being in the
\textit{l}eft or \textit{r}ight TLR denotes the logic 0 or 1 state
\cite{ref:OQC} for a single qubit. This is analogous to the
conventional optical cavity mode representation of photonic qubit
where the information is encoded by which cavity the photon is in
\cite{ref:text}. Notice for a dilution refrigerator temperature of
$40$mK, the thermal photon number in the TLRs is smaller than
$10^{-10}$ and thus the 0 or 1 photon state for the TLRs is an
excellent approximation. To effect arbitrary transformations on this
single qubit, we need to be able to shift the relative energies of the
TLRs and transfer photons between them, which implement the
functionalities of phase shifters and beam splitters in optics.  We
realize this by coupling the TLRs capacitively to current biased
Josephson junctions (CBJJ) as shown in Fig. \ref{fig1} (a). As the
simplest Josephson qubit, CBJJ has the advantage that its level
splitting can be easily adjusted by the bias current. Approximating
the CBJJs as two-state systems with adjustable energy splittings
$\Omega_c$ and $\Omega_r$ \cite{ref:two-level}, we can write the
system Hamiltonian $H= \hbar\omega_0 (\hat{a}^{\dagger} \hat{a} +
\hat{b}^{\dagger} \hat{b}) + \frac{1}{2} \hbar \Omega_c \sigma^z_c
+\frac{1}{2} \hbar \Omega_r \sigma^z_r + \hbar g_c [(\hat{a} +\hat{b})
\sigma^+_c +(\hat{a}^{\dagger} +\hat{b}^{\dagger}) \sigma^-_c] + \hbar
g_r (\hat{b} \sigma^+_r + \hat{b}^{\dagger} \sigma^-_r)$
\cite{ref:coupstr}, where $\sigma^{z,\pm} _{c,r}$ are the Pauli
matrices of the coupling and right CBJJ, $\hat{a}$, $\hat{b}$ are the
annihilation operators for photons in the two TLRs, and the coupling
strengths $g_{c,r}=\omega_0C_{c,r}/\sqrt{2C(C_J^{c,r}+2C_{c,r})}$,
$C_J^{c,r}$ the capacitance of the coupling and right CBJJ.

Since the CBJJ energies can be easily adjusted by tuning the bias
current, we can control the interactions between the TLRs and CBJJs.
To transfer photons between the TLRs, we adjust the bias currents of
the CBJJs to tune $\Omega_r$ faraway from $\omega_0$ so the right CBJJ
has no effect. We further tune the coupling CBJJ close to resonance
with $\omega_0$ and work in the dispersive region where the magnitude
of detuning $\Delta_{c} = \Omega_c-\omega_0$ is much greater than
$g_c$. 
Assuming the CBJJ was prepared in the ground
state, its virtual excitation gives rise to the following effective
Hamiltonian for the TLRs \cite{ref:disp} in the rotating frame defined
by the uncoupled TLR Hamiltonian:
\begin{equation}
  H_{eff}= \frac{\hbar g_c^2}{\Delta_c}(\hat{a}\hat{b}^{\dagger} +\hat{a}^{\dagger}\hat{b}) + \frac{\hbar g_c^2}{\Delta_c}(\hat{a}^{\dagger}\hat{a} +\hat{b}^{\dagger}\hat{b}).
\label{eq:Heff}
\end{equation}
Since there is only 1 photon in the system, $\hat{a}^{\dagger}\hat{a}
+\hat{b}^{\dagger}\hat{b} =1$, energy shifts described in the second
term in $H_{eff}$ is a constant.
The first exchange term implements the photon transfer operation. A
photon can be transferred between the two TLRs with a rate
$\frac{g_c^2}{2\pi\Delta_c}$, which is about $20$MHz for $C_J^c =
0.5$pF, $C_c = 23$fF, and $\Delta_c/2\pi = 2$GHz \cite{ref:para1,
  ref:para2}. 

To shift the relative energies of the TLRs, we tune the coupling
CBJJ far off resonance and tune the right CBJJ into the dispersive
region. Similarly, an effective Hamiltonian $(\hbar g_r^2/\Delta_r)
\hat{b}^ {\dagger}\hat{b}$ results which gives a relative phase when
the photon is in the right TLR.

We need to study the decoherence properties of our scheme to analyze
its reliability. The photonic qubits have superb coherence and their
life times are orders of magnitude longer than that of superconducting
qubits. For TLRs fabricated on superconducting chips, a high quality
factor of $10^6 - 10^7$ has been demonstrated \cite{ref:highQ}. For
TLR frequencies of tens of GHz, the photon loss rate $\kappa/2\pi$ can
be as low as KHz. 
In contrast, the CBJJ has a short decoherence time, and we assume its
dephasing rate $\Gamma_2/2\pi \approx $1MHz. The CBJJ's decay rate
from the excited state $\Gamma_1/2\pi$ is on the order of $0.1$MHz.

A major advantage of our scheme is that the relatively lossy CBJJ does
not damp the coherence of the photonic qubits much since it is only
virtually excited. The CBJJ's decay from the virtually excited state
increases the photon's loss rate by
$(g_{c,r}/\Delta_{c,r})^2\Gamma_1$, which is not a concern since
$(g_{c,r}/\Delta_{c,r})^2\Gamma_1$ is no greater than $\kappa$. To
study the effect of the CBJJ's dephasing rate $\Gamma_2$, we model the
dephasing effect as the result of a random fluctuation $\delta_n$ in
the CBJJ's energy splitting. This introduces an uncertainty in the
detuning during for instance a photon transfer operation, $\Delta_c =
\Omega_c-\omega_0\rightarrow \Delta_c +\delta_n$. Therefore, the
system will have a random Hamiltonian $H_{noise}=-\hbar
(g_c/\Delta_c)^2\delta_n (\hat{a}^{\dagger} \hat{b} +\hat{a}
\hat{b}^{\dagger})$ in addition to that in Eq.
(\ref{eq:Heff}). Assuming the distribution of $\delta_n$ is Gaussian,
we can estimate the photonic qubit's decoherence time due to
$H_{noise}$ by using the free induction decay function
\cite{ref:deph}. Since the free inductor decay function is determined
by the spectral density of $\delta_n$, which in turn is related to the
CBJJ's dephasing rate $\Gamma_2$, it can be estimated that the
system's dephasing rate is no greater than $2(\frac{g_c^2}{\Delta_c^2}
)\Gamma_2$.

Following the quantum theory of damping, we now calculate the gate
error of a photon transfer operation under the influence of cavity
loss and CBJJ dephasing.  using the Master equation for the qubit's
density matrix $\rho$, $d\rho/dt =-i[H_{eff}, \rho]
+\kappa[\hat{a}\rho \hat{a}^{\dagger} -\frac{1}{2} \hat{a}^{\dagger}
\hat{a}\rho -\frac{1}{2} \rho\hat{a}^{\dagger} \hat{a}]
+\kappa[\hat{b}\rho \hat{b}^{\dagger} -\frac{1}{2} \hat{b}^{\dagger}
\hat{b}\rho -\frac{1}{2} \rho\hat{b}^{\dagger} \hat{b}]
+2(\frac{g_c}{\Delta_c})^2 \Gamma_2 [(\hat{a}^{\dagger} \hat{b}
+\hat{a} \hat{b}^{\dagger}) \rho (\hat{a}^{\dagger} \hat{b} +\hat{a}
\hat{b}^{\dagger}) -\rho]$. The gate error probability of a single
qubit bit flip is plotted in Fig. \ref{fig1}(b) as a function of
$\kappa$ and $\Gamma_2$. The result indicates that, for already
demonstrated $\Gamma_2/2\pi =1$MHz and $\kappa/2\pi =10$kHz
\cite{ref:cbjjdeph}, the gate error is on the order of $10^{-3}$.

\begin{figure}[h]
\subfigure []
{\label{2a}\includegraphics[width=0.4\columnwidth]{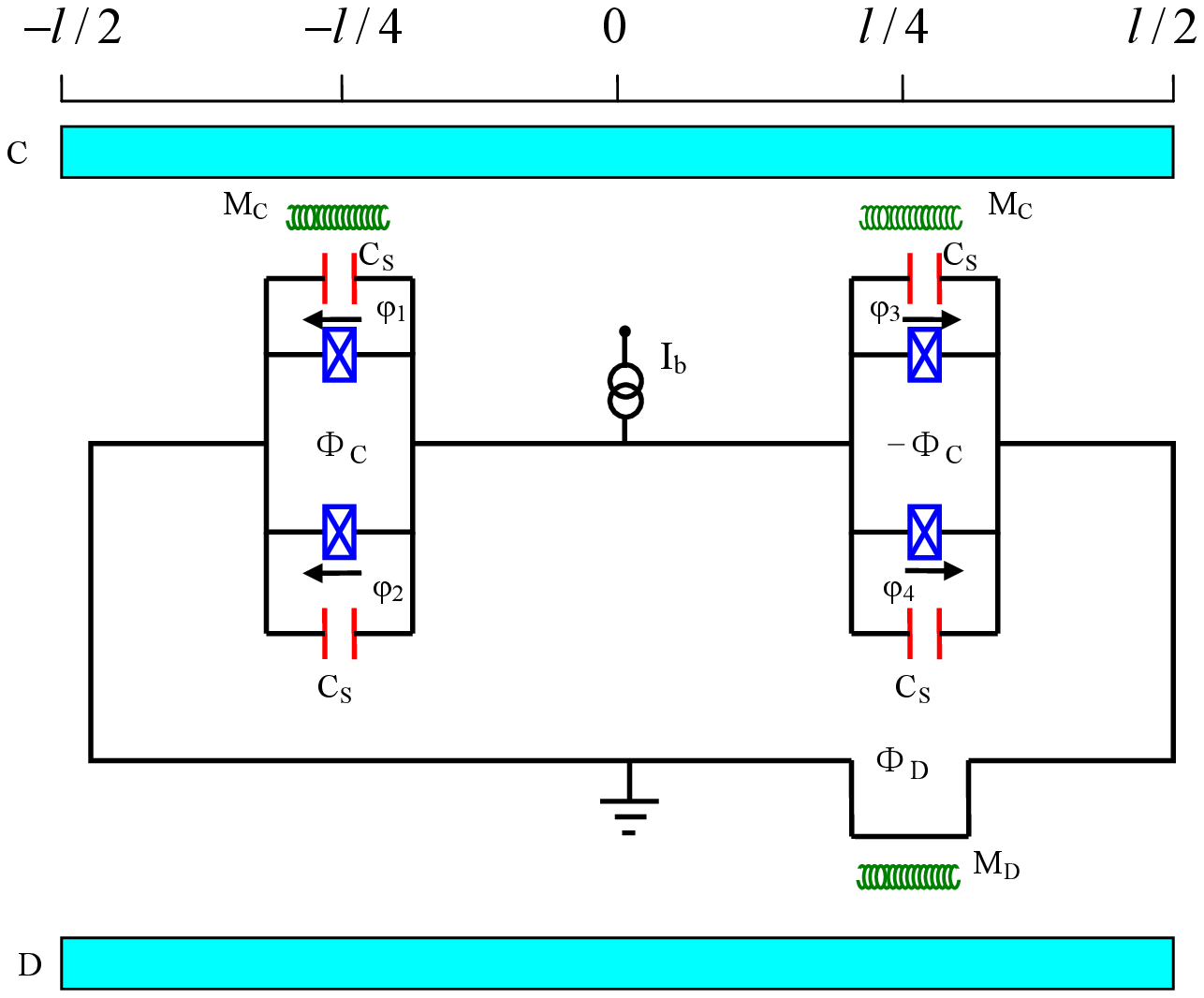}}
\subfigure[]{\label{2b}\includegraphics[width=0.3\columnwidth]{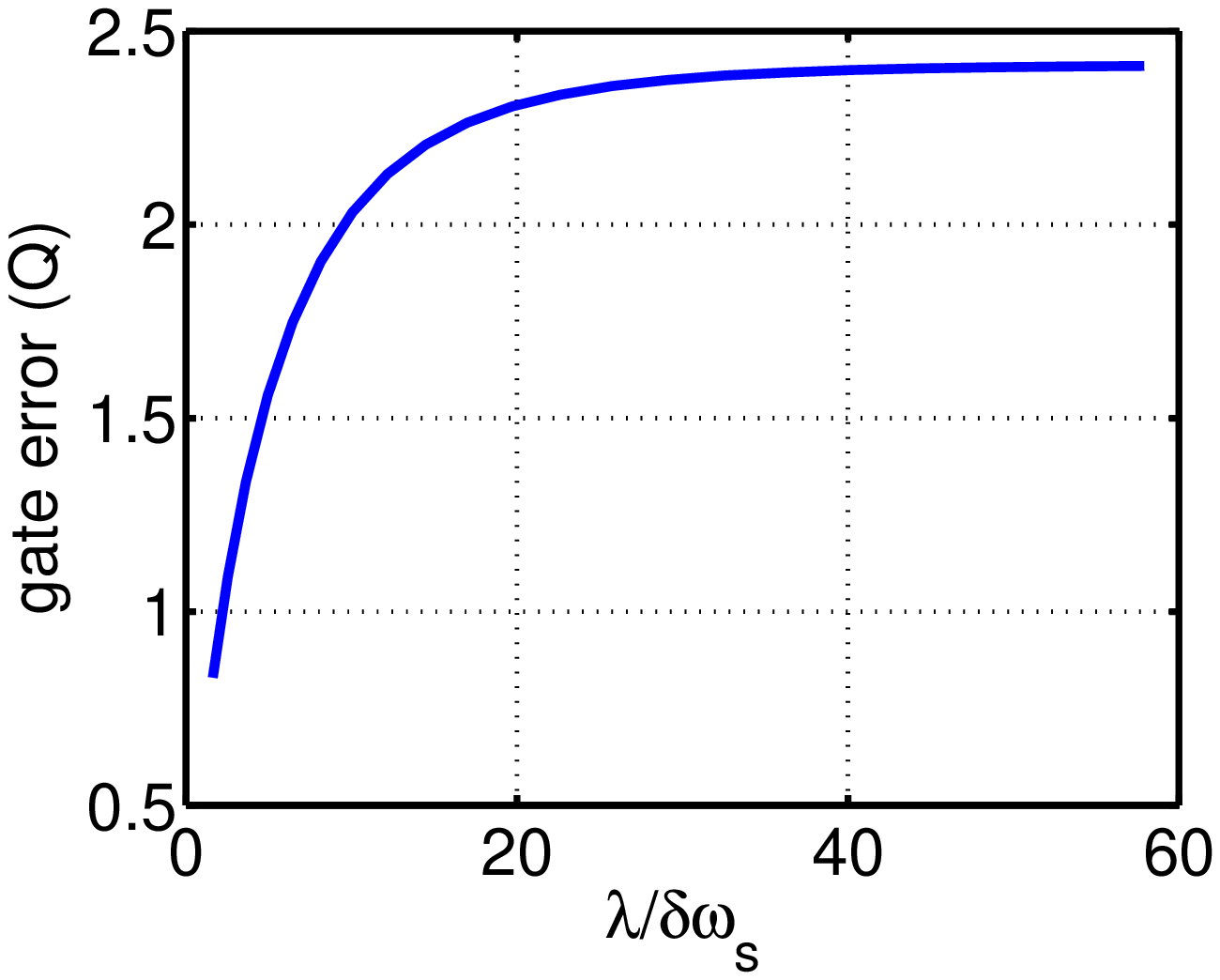}}\\%
\subfigure[]{\label{2c}\includegraphics[width=0.8\columnwidth]{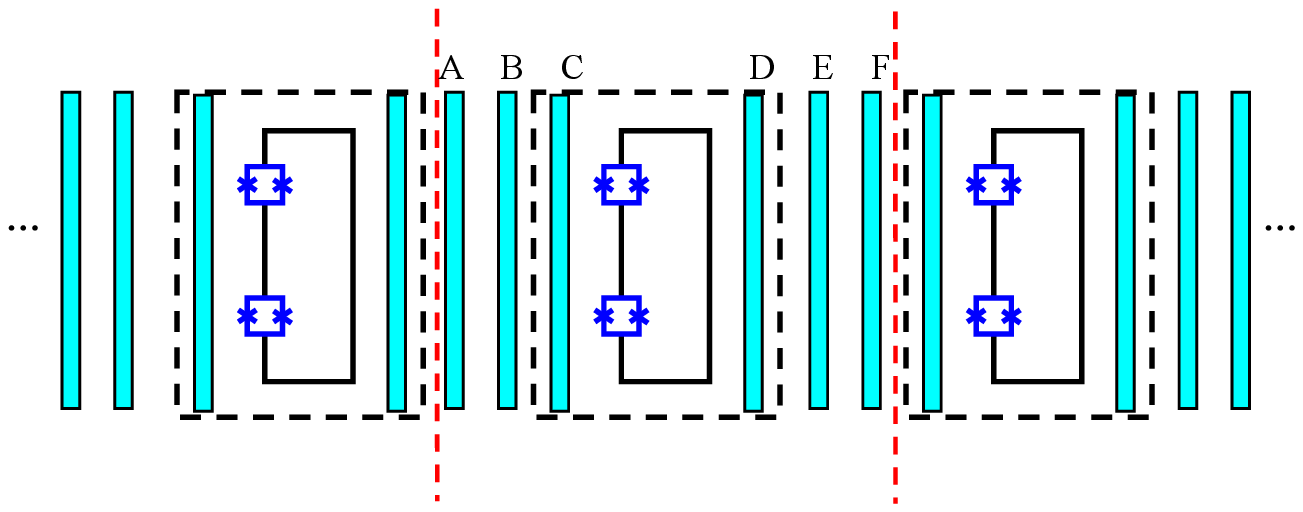}}
\caption{(a) The 4-junction SQUID to interact photons. 
   (b) Dependence of the error probability ($10^{-Q}$) of a
  controlled phase gate on the ratio between photon transfer rate
  $\lambda$ and uncertainty in photon energy shift $\omega_s$. (c) The
  setup for controlled photon interaction. The circuit in the
  dashed-line box is that in (a). The TLRs are coupled to CBJJs (not
  shown) so that single bit operations including photon transfer can
  be performed. 
} \label{fig2}
\end{figure}

The manipulations demonstrated so far perform linear optics. We still
need a mechanism to induce interactions between photons. This is a
major difficulty in conventional optics. However, at microwave
frequencies, we can take advantage of the strong nonlinearities in
Josephson devices to interact photons.

We consider the low current biased 4-junction SQUID (FJS) device
\cite{ref:nems} in Fig. \ref{fig2}(a). The two small identical SQUIDs
are coupled inductively to TLR $C$ of length $l$ at positions $\pm
l/4$.  (This does not mean that the FJS must extend to a length of
$l/2$ because the TLR can be layed out in a zig-zag fashion.)  Since
$l$ is much larger than the dimension of the FJS, we can adopt the
long wave approximation and use the TLR current at the SQUIDs'
locations in calculating the SQUIDs' flux bias. At the two coupling
points, the TLR currents are the largest in magnitude and opposite in
direction. The main loop is coupled to TLR $D$ at $l/4$.

Assuming there is no other external flux bias, the small SQUIDs and
big loop are biased by the TLR currents $I_C= \mp iI_{C0}(\hat{c}
-\hat{c}^{\dagger})$ and $I_D= -iI_{D0}(\hat{d}
-\hat{d}^{\dagger})$, where $I_{C0}= \sqrt{\hbar\omega_c/L_c}$ and
$I_{D0}= \sqrt{\hbar\omega_d/L_d}$ the zero point current
fluctuations in the TLRs, $L_{c,d}$ the inductance of the TLRs and
$\hat{c}$, $\hat{d}$ the annihilation operators for photons in $C$
and $D$. 
We can work out the system's Hamiltonian, $H= H_{TLR} +
H_{FJS} +H_{int}$, where $H_{TLR} =\hbar (\omega_c +\omega^c_s)
\hat{c}^{\dagger}\hat{c} +\hbar (\omega_d
+\omega^d_s)\hat{d}^{\dagger}\hat{d}$, $H_{FJS} = -\frac{E_C}{2}
\frac{\partial ^2} {\partial\phi^2} -\frac{\hbar
  I_b\phi}{2e} -4E_J^0\cos\phi$, and
\begin{equation}
  H_{int}= \hbar \omega_{int}\hat{c}^{\dagger}\hat{c} \hat{d}^{\dagger}\hat{d} = \hbar \omega_{int}\hat{n}_c\hat{n}_d.
\label{eq:int}
\end{equation}
In these equations, $E_c=(2e)^2/4(C_J+C_s)$ and $E_J = \hbar I_c/2e$
are the charging energy and Josephson energy of the junctions, where
$C_J$ and $I_c$ are the junctions' capacitance and critical current.
$\phi$ is the average phase of the 4 junctions determined by the low
bias current $I_b \approx 0$ of the FJS. The frequency shift
$\omega^c_s = -2E_J^0(\phi^2\chi_c^2+\chi_c^2\chi_d^2)/\hbar$,
$\omega_s^d = -2E_J^0(\phi^2\chi_d^2+\chi_c^2\chi_d^2)/\hbar$ where
$\chi_{c} =\pi M_{C}I_{C0}/(\pi L_sI_c+\Phi_0)$, $\chi_{d} = \pi
M_DI_{D0}/[\pi(L_s+L_L)I_c+\Phi_0]$, $L_s$ and $L_L$ are the
self-inductances of the small SQUIDs and the circuit loop. To simplify
the expressions, we set $\chi = \chi_c = \chi_d$ and denote the photon
frequency shift $\omega_s \equiv \omega_s^c = \omega_s^d$. The photon
interaction strength $\omega_{int} = -4E_J^0\chi^4\cos\phi/\hbar$. In
deriving the system Hamiltonian, we have used the rotating wave
approximation and 
dropped terms that will be oscillating fast in the rotating frame
defined by $H_{TLR}$. We have also dropped terms involving creation
and annihilation of two photons. These terms have no effect since
there is no more than 1 photon in the TLRs in our scheme.

We operate with a low bias current $I_b \approx 0$ for the FJS so that
$\langle cos{\phi} \rangle$ is large and the FJS' energy splitting is
far away from the frequencies of the TLR photons. Thus, the FJS will
not be excited by the TLR photons and they hardly get entangled. The
FJS then acts as a ``nonlinear medium" and Eq. (\ref{eq:int})
describes the interaction between photons in $C$ and $D$ modulated by
the FJS' phase. For $I_c = 50\mu$A, $L_s \approx 10$pH, and $M_C
\approx 80$pH \cite{ref:para3}, the photon interaction strength
$\omega_{int} \approx 1$MHz, much greater than the photon loss
rate. Unfortunately, there are difficulties in using this interaction
for quantum computing. 
First, $\phi$ has fluctuations in it due to the FJS' charging energy
and thus the interaction strength is not a constant. Also, it is not
easy to turn off the interaction. Tuning $\phi$ close to $\pi/2$
requires biasing the FJS close to its critical current which makes the
system unstable. The uncertainty in $\phi$ grows too.

To have the photons interact only when needed, we use a setup shown
in Fig. \ref{fig2} (c). Here TLRs $A$, $B$ and $E$, $F$ are two
qubits with photons being in $A$ and $F$ representing their logic 0
state. When both qubits are in the 1 state, we can use the photon
transfer operation discussed earlier to transfer the photons from
$B$ and $E$ to the auxiliary TLRs $C$ and $D$ whose frequencies are
made different than that of the qubit TLRs by $\omega_s$ to account
for their energy shifts. Once the photons are in $C$ and $D$ they
can interact due to coupling to the FJS. Afterwards, we transfer
them back to $B$ and $E$.

To stabilize the FJS' phase, we shunt its junctions with large
capacitances $C_s$ as shown in Fig. \ref{fig2}(a). At low bias
currents the FJS' behavior can be very well approximated by that of a
harmonic oscillator and the distribution of $\phi$ is given by its
ground state wave function $\sqrt{\alpha/\sqrt{\pi}} \exp[-\alpha
^2(\phi- \phi_0)^2/2]$, where $\alpha =\sqrt[4]{4E_J^0\cos\phi_0/E_c}$
and $\phi_0 = \langle\phi\rangle = \arcsin{\frac {\hbar
    I_b}{8eE_J^0}}$.  If we choose a total capacitance $C_J+C_s =
20$pF, the relative uncertainty $\delta (\omega_{int})/\omega_{int}
\approx 10^{-4}$. Such a small error is not a concern for the photon
interaction term. However uncertainties in the photon energy shift
terms $\hbar\omega_s$ can be comparable to $\hbar\omega_{int}$ and can
cause large errors.

We employ a two-phase technique in the spirit of spin-echo to address
this problem. In phase 1, we first do a photon transfer operation
between $B$, $C$ and $E$, $D$ with a speed relatively fast compared to
$\omega_{int}$ and $\delta\omega_s =
-2E_J^0\chi^2\delta(\phi^2)/\hbar$, the uncertainty in the photon
frequency shift. We then wait for a desired interaction time $t=
\pi/\omega_{int}$ after which we do another photon transfer between
$B$, $C$ and $E$, $D$. In phase 2, we first perform a bit flip for the
2 qubits, in other words do a photon transfer operation between $A$,
$B$ and $E$, $F$. We then repeat phase 1. At the end, we perform a bit
flip on the two qubits again. In this process, depending on their
initial states the qubits will acquire the same random phase due to
$\delta\omega_s$ in either phase 1 or 2, thus removing the effect of
the randomness in the photon energy shifts.  The end result is a $\pi$
phase shift on the 2-qubit states if they are both in 0 or 1
initially. This is equivalent to a controlled phase gate and it
enables universal quantum computing in combination with the single
qubit operations.

If the photon transfer operation between $B$, $C$ and $E$, $D$ was
perfect, the controlled phase gate would be exact. 
phase shift in $C$ and $D$ could be eliminated completely.  However,
since the photons in $C$ and $D$ will interact with the FJS even
during the photon transfer, our control phase gate will have errors.
This can be seen by examining the system Hamiltonian during the photon
transfer (in the rotating frame) $H= \hbar
\lambda_{bc}(\hat{b}^{\dagger}\hat{c} + \hat{b}\hat{c}^{\dagger}) +
\hbar \lambda_{de}(\hat{d}^{\dagger}\hat{e} +
\hat{d}\hat{e}^{\dagger}) -2E_J^0\chi^2\phi^2(\hat{c}^{\dagger}\hat{c}
+\hat{d}^{\dagger}\hat{d}) -\hbar\omega_{int} \hat{c}^{\dagger}\hat{c}
\hat{d}^{\dagger}\hat{d}$. The first two terms are used for the photon
transfer operation, however the remaining terms cannot be turned off
making the photon transfer imperfect.  Obviously, the fidelity of our
controlled phase gate will be improved by making the photon transfer
frequency $\lambda_{bc}$ and $\lambda_{de}$ large compared to
$\delta\omega_s$ and $\omega_{int}$. We numerically studied our
control phase gate using the full Hamiltonian and plotted the gate
error in Fig \ref{fig2} (b). We set $\lambda \equiv \lambda_{bc} =
\lambda_{de}$. For our choice of system parameters,
$\lambda/\delta\omega_s \approx 20$ and the gate error is on the order
of $10^{-3}$.

Our microelectronic system is easily scalable 
as shown in Fig. \ref{fig2} (c). We can extend the setup for
the control phase gate in both ends to integrate many TLR qubits on
the same chip with an FJS between each pair of qubits. This is a 1d
architecture with controllable interactions between adjacent qubits
that can be scaled to a large number of qubits.

\begin{figure}[h]
\subfigure []
{\label{3a}\includegraphics[width=0.4\columnwidth]{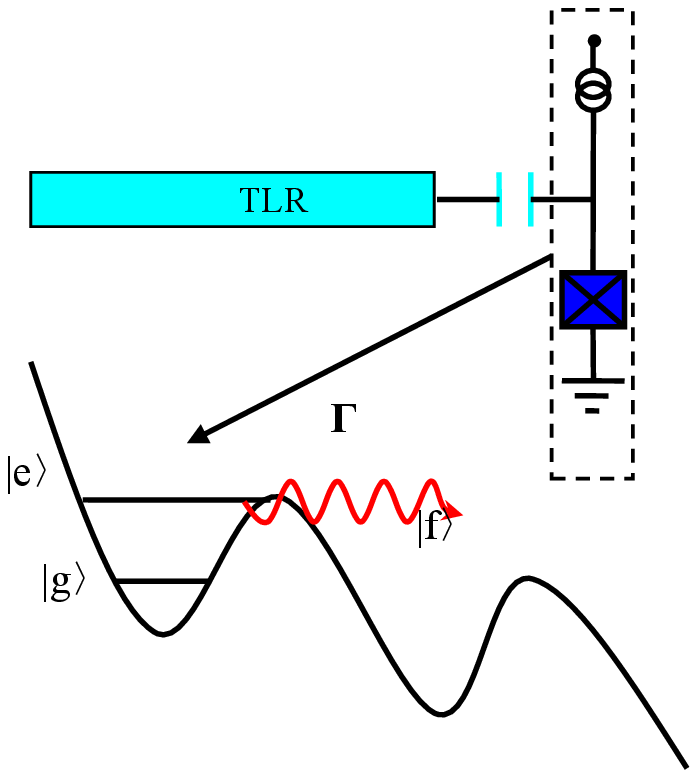}}%
\subfigure[]{\label{3b}\includegraphics[width=0.4\columnwidth]{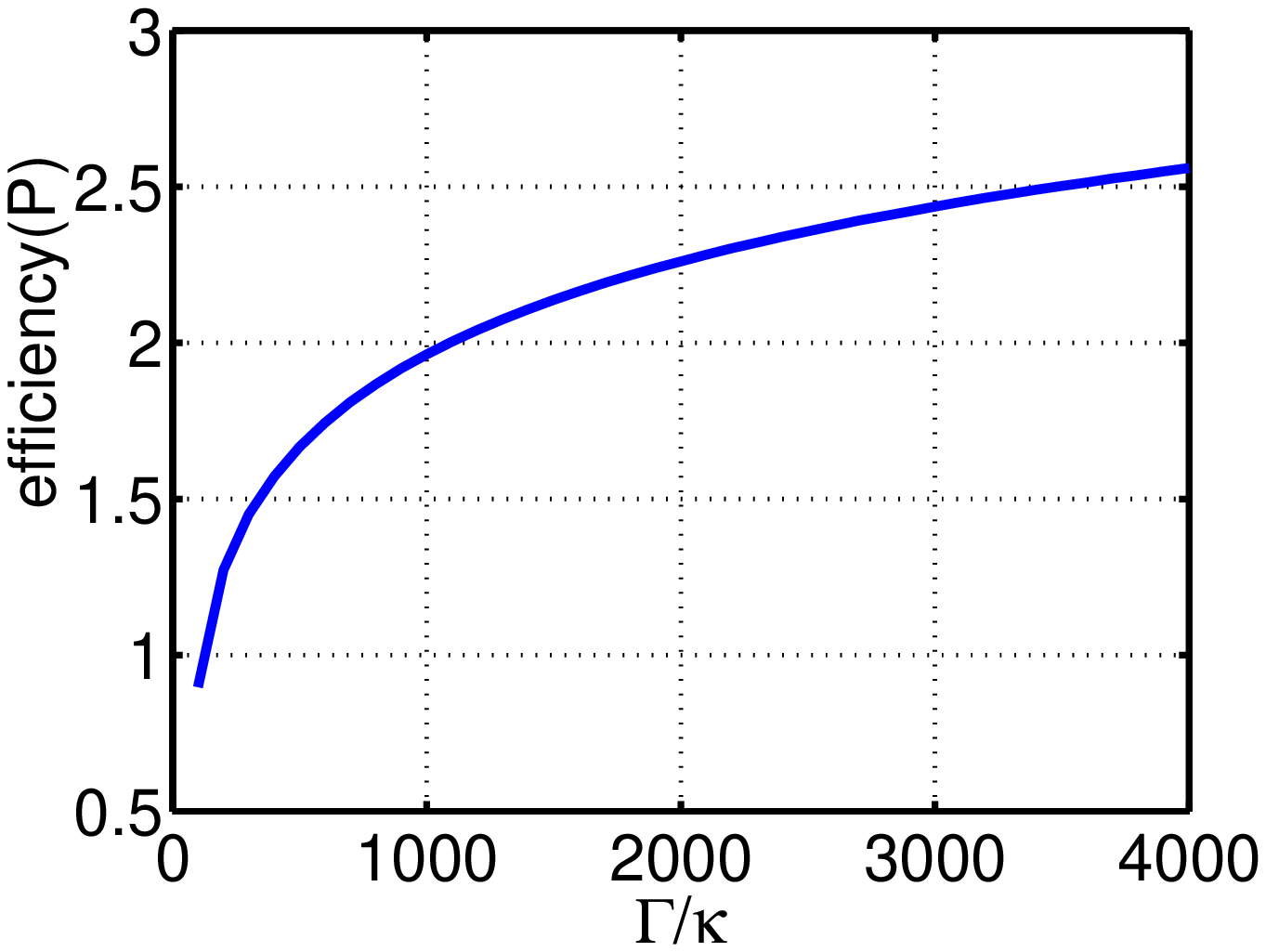}}
\caption{(a) The photon detection scheme. (b) The dependence of
detector efficiency($1-10^{-P}$) on the ratio between the escape
rate $\Gamma$ and photon loss rate $\kappa$. The coupling strength
$g_{td}/2\pi = 100$MHz, photon loss rate $\kappa/2\pi = 10$kHz, CBJJ
decay and dephasing rate $\gamma_T/2\pi = 100$kHz,
$\gamma_\varphi/2\pi = 1$MHz. } \label{fig3}
\end{figure}

In order to perform photonic qubit quantum computing, we still need to
be able to generate and detect single photons. Photon generation on
superconducting chip has been demonstrated experimentally
\cite{ref:PGE1, ref:PGE2, ref:PGE3}. For photon detection
\cite{ref:detector, ref:qnd}, we again consider a CBJJ coupled to a
TLR of frequency $\omega_0$ as shown in Fig. \ref{fig3} (a). The CBJJ
is prepared in the ground state $|g \rangle$ in the well of its
washboard potential. We also make use of an unstable excited state $|e
\rangle$ where the CBJJ can tunnel to the voltage state with a large
rate $\Gamma$. By adjusting the CBJJ's bias current, we can tune the
CBJJ in resonance with $\omega_0$. The CBJJ will then be excited by
the TLR photon to $|e \rangle$. When it escapes from $|e \rangle$, a
detectable voltage appears across the CBJJ.

Though an easy and reliable method, our scheme may fail to detect a
photon in the TLR due to the photon decay and the CBJJ's intra-well
decay and decoherence.  The TLR photon may decay before being detected
by the CBJJ. The CBJJ's intra-well decay from $|e \rangle$ to $|g
\rangle$ and its finite decoherence time are concerns too.  To study
the influence of the photon loss rate and CBJJ's intra-well decay and
decoherence on the efficiency of our photon detector, we model it as a
3-state system shown in Fig. \ref{fig3} (a), where $|f \rangle$
represents the voltage state. We use the Master equation $d\rho/dt
=-i[H,\rho] +\mathcal{L} \rho$. Here, $\rho$ is the density matrix of
the system, the system Hamiltonian $H= \delta \hat{a}^{\dagger}\hat{a}
+g_{td}(\hat{a}^{\dagger} \sigma_{ge} +\hat{a} \sigma_{eg})$, the
detuning $\delta =\omega_0 -\mu$, $\mu$ the frequency difference
between $|e \rangle$ and $|g \rangle$. The Liouvillian $\mathcal{L}
\rho =\frac{\kappa}{2} \mathcal{L}[\hat{a}]\rho
+\frac{\Gamma}{2}\mathcal{L}[\sigma_{ef}]\rho +\frac{\gamma
  _{T}}{2}\mathcal{L}[\sigma_{eg}]\rho
+\frac{\gamma_\varphi}{2}\Sigma_{i = g,e}\mathcal{L}[|i\rangle\langle
i|]\rho$, where $\mathcal{L}[\hat{O}]\rho \equiv
2\hat{O}\rho\hat{O}^\dagger -
\hat{O}^\dagger\hat{O}\rho-\rho\hat{O}^\dagger\hat{O}$. $\kappa$ is
the decay rate of the photon in the TLR, $\gamma_T$ and
$\gamma_{\varphi}$ are the intra-well decay rate and dephasing rate of
the CBJJ, $\sigma_{ij} =|i\rangle \langle j|$ for $i,j=g,e,f$, and
$\sigma_z= |e\rangle \langle e| -|g\rangle \langle g|$.

Assuming initially there is a photon in the TLR and the CBJJ is in $|g
\rangle$, we plot the detecting efficiency (the probability the CBJJ
ends up in $|f \rangle$) in Fig. \ref{fig3} (b) as a function of
$\Gamma/\kappa$. As can be seen, the efficiency is high even for
moderately large escaping rate $\Gamma$. For $\Gamma/2\pi =20$MHz
\cite{ref:Gamma} and $\kappa/2\pi = 10$kHz, the detection efficiency
is above $99\%$. Also, it is demonstrated in our simulation that the
influence of $\gamma_{\varphi}$ to the detection efficiency is minor
and thus the detecting CBJJ does not need to have long decoherence
times.

In summary, we have shown that, by using TLR microwave photons as
qubits and Josephson devices as optical devices, a superconducting
chip provides an ideal implementation for fully integrated photonic
qubit quantum computing. Thanks to our careful design, high gate
fidelities can be achieved and thus our scheme is a realistic
approach. Since our system is based on existing mature technologies,
fast experimental progress can be expected to bring integrated
photonic qubit quantum computing to reality. The novel idea of using
on-chip microwave photons as qubits also opens the possibility of
investigating many interesting optical quantum effects in an
integrated system.

This work was supported by NNSF of China (Grant Nos. 10875110,
60621064, 10874170) and National Fundamental Research Program of China
(No. 2006CB921900).


\begin{references}

\bibitem{ref:Integrated_OQC} A. Politi, M. J. Cryan, J. G. Rarity,
  S. Yu, and J. L.O'Brien, Science {\bf 320}, 646 (2008).

\bibitem{ref:onchip} R. G. Hunsperger, {\it Integrated Optics}, Springer,
2002.

\bibitem{ref:CJ} F. Plastina and G. Falci, Phys. Rev. B {\bf 67},
224514 (2003).

\bibitem{ref:disp} X. Zhou {\it et al.}, Phys. Rev. A {\bf 69},
030301(R) (2004).

\bibitem{ref:tlr} A. Blais, R. S. Huang, A. Wallraff, S. M. Girvin, and R.
J. Schoelkopf, Phys. Rev. A. {\bf 69}, 062320 (2004).

\bibitem{ref:OQC} I. L. Chuang and Y. Yamamoto, Phys. Rev. A {\bf
52}, 3489 (1995).

\bibitem{ref:text} M. A. Nielsen and I. L.Chuang, ``\textit{Quantum
    computation and quantum information},'' Cambridge University
  Press, 2000.

\bibitem{ref:two-level} Notice that even though we use the 2-state
  formalism for the CBJJ for the convenience of keeping only the
  relevant terms in the system Hamiltonian, our scheme does not rely
  on treating the CBJJ as a 2-state system. The CBJJ will only be
  virtually excited so potential problems with the 2-state
  approximation such as insufficient anharmonicity do not affect our
  results at all.

\bibitem{ref:coupstr} Y. Hu \textit{et al.}, 
Phys. Rev. A {\bf 75}, 012314 (2007).

\bibitem{ref:para1} A. Wallraf \textit{et al.}, 
  Nature (London) {\bf 431}, 162 (2004).

\bibitem{ref:para2} M. A. Sillanpaa, J. I. Park, and R. W. Simmonds,
Nature (London) {\bf 448}, 438 (2007).

\bibitem{ref:highQ} P. K. Day \textit{et al.}, 
Nature (London) {\bf 425}, 817 (2003).


\bibitem{ref:deph} G. Ithier \textit{et al.}, 
Phys. Rev. B {\bf 72}, 134519 (2005).

\bibitem{ref:cbjjdeph} J. M. Martinis, Quantum Inf. Process. {\bf 8}, 81(2009).

\bibitem{ref:nems} X. Zhou and A. Mizel, Phys. Rev. Lett. {\bf 97},
  267201 (2006).

\bibitem{ref:para3} S. H. W. van der Ploeg {\it et al.}, Phys. Rev.
Lett. {\bf 98}, 057004 (2007).

\bibitem{ref:PGE1}B. T. H. Varcoe, S. Brattke, M. Weidinger and H. Walther, Nature (London) {\bf 403}, 743
(2000).

\bibitem{ref:PGE2}A. A. Houck \textit{et al.}, 
  Nature (London) {\bf 449} 328 (2007).

\bibitem{ref:PGE3} M. Hofheinz \textit{et al.}, 
  Nature (London) {\bf 454}, 310 (2008).

\bibitem{ref:detector} G. Romero, J. J. Garcia-Ripoll, and E. Solano,
  Phys. Rev. Lett. {\bf 102}, 173602 (2009).

\bibitem{ref:qnd} F. Helmer, M. Mariantoni, E. Solano, and
  F. Marquardt, Phys. Rev. A {\bf 79}, 052115 (2009).

\bibitem{ref:Gamma} K. B. Cooper \textit{et al.}, 
  Phys. Rev. Lett. {\bf 93}, 180401 (2004).

\end{references}
\end{document}